\documentclass[preprint,eqsecnum,aps,prd,nofootinbib]{revtex4}

\usepackage[pdftex]{color,graphicx}
\usepackage{enumerate}
\usepackage{amsmath}
\usepackage{amsfonts}
\usepackage{verbatim}

\def\be{\begin{equation}}
\def\bea{\begin{eqnarray}}
\def\ee{\end{equation}}
\def\eea{\end{eqnarray}}
\usepackage{hyperref}

\begin{document}

\title{Two-dimensional semiclassical static black holes: Finite-mass correction to the Hawking temperature and outflux}

\author{Adam Levi and Amos Ori}


\affiliation{ \; \; \; \; \;   \\  Department of Physics \\ Technion-Israel Institute of Technology \\ Haifa 3200, Israel \\  }

\begin{abstract}
In the two-dimensional framework, the surface gravity of a (classical) black hole is independent of its mass $M$. As a consequence, the Hawking temperature and outflux are also independent of $M$ at the large-$M$ limit. (This contrasts with the four-dimensional framework, in which the surface gravity and temperature scale as $1/M$.) However, when the semiclassical backreaction effects on the black-hole geometry are taken into account, the surface gravity is no longer $M$-independent, and the same applies to the Hawking temperature and outflux. This effect, which vanishes at the large-$M$ limit, increases with decreasing $M$. Here we analyze the semiclassical field equations for a two-dimensional static black hole, and calculate the leading-order backreaction effect ($\propto 1/M$) on the Hawking temperature and outflux. We then confirm our analytical result by numerically integrating the semiclassical field equations.
\end{abstract}

\date{\today}

 \maketitle

\section{Introduction}

In classical General Relativity, a black hole (BH) is absolutely black:
It does not emit any radiation. The situation changes, however, when
quantum effects are taken into account. The semiclassical extension
of General Relativity considers quantum fields which live on the background
of a well-defined classical geometry (e.g. a black-hole). Within this
framework it was found \cite{Hawking} that a black hole actually
has a finite temperature, the \emph{Hawking temperature} $T_{H}$.
Accordingly the BH emits thermal radiation, and evaporates within
a finite time.

Hawking's analysis \cite{Hawking} revealed that the temperature of
the semiclassical BH is $T_{H}=\kappa/2\pi$, where $\kappa$ is the
\emph{surface gravity} of the BH. Throughout this paper we use General-Relativistic
units $c=G=1$ (and the same for the Boltzmann constant), and we also
set $\hbar=1$ (following Ref. \cite{CGHS}).
\footnote{In four dimensions (and in fact in any $d\ne 2$ dimensions) setting $c=G=\hbar=1$ merely amounts to a choice of units. However in two dimensions this is not the case, because $c^{-3} G \hbar$ becomes dimensionless. Thus setting $\hbar=1$ here is an arbitrary choice.}
The temperature $T_{H}$
is thus uniquely determined by the background BH geometry. For a 4-dimensional
(4D) Schwarzschild BH of mass $M$ the surface gravity is $\kappa=1/(4M)$,
hence $T_{H}=1/(8\pi M)$.

In the semiclassical theory, the quantum field yields an Energy-momentum
contribution $\hat{T}_{\alpha\beta}$, to be inserted at the right-hand
side of the semiclassical Einstein equation $G_{\alpha\beta}=8\pi\hat{T}_{\alpha\beta}$.
This \emph{renormalized stress-energy tensor} $\hat{T}_{\alpha\beta}$
is a tensor field in spacetime, which depends on the spacetime geometry
(as well as on the quantum state of the matter field under consideration).
The Hawking radiation is the most obvious manifestation of this semiclassical
$\hat{T}_{\alpha\beta}$: It is the outging component of $\hat{T}_{\alpha\beta}$,
evaluated at future null infinity (FNI). But obviously $\hat{T}_{\alpha\beta}$
includes other components as well, and is also position-dependent.
For example, an evaporating BH must be endowed with a negative ingoing
component of $\hat{T}_{\alpha\beta}(x^{\mu})$ at the horizon: It
is this ingoing component which is directly responsible for the steady
decrease of the BH mass.

The classical Schwarzschild solution is a vacuum solution of the Einstein
equation. Obviously, the semiclassical contribution $\hat{T}_{\alpha\beta}$
must modify the BH geometry, which will no longer be a pure vacuum
solution. Instead, the BH metric $g_{\alpha\beta}$ is to be determined
now from the semiclassical Einstein equation $G_{\alpha\beta}=8\pi\hat{T}_{\alpha\beta}$.

Since in a semiclassical BH $g_{\alpha\beta}$ is no longer the Schwarzschild
geometry, the BH's surface gravity will deviate from its classical
value $\kappa=1/(4M)$. This deviation is small for a macroscpic BH
(namely, $M\gg m_{p}$, where $m_{p}$ denotes the Planck mass), and
is expected to decrease $ $with increasing $M$. This change in $\kappa$
yields a corresponding change in the BH temperature and outflux. The
main objective of this paper is to explore this (mass-dependent) change
in the temperature and outflux of a semiclassical BH, caused by the
deviation of the background geometry from the classical one. (Though,
we shall actually tackle this problem in two rather than four spacetime
dimensions, for reasons which we shortly explian.)

Presently there is no known explicit expression for the renormalized
stress-energy tensor $\hat{T}_{\alpha\beta}$ in 4D (even for spherically-symmetric
spacetimes). This makes it hard to construct the semiclassical BH
geometry and evaluate its surface gravity. Fortunately
the situation is much simpler in the two-dimensional (2D) framework,
wherein $\hat{T}_{\alpha\beta}$ is explicitly known for a generic
background metric. This motivates us to address
this issue---the mass-dependent semiclassical correction to the Hawking
temperature and outflux---in the 2D framework.

About two decades ago Callan, Giddings, Harvey and Strominger (CGHS)
\cite{CGHS} introduced a formalism of 2D dilaton gravity in which
the metric is coupled to a dilaton field $\phi$ and to a large number
$N$ of identical massless scalar fields. In this 2D framework $\hat{T}_{\alpha\beta}$
is known explicitly, allowing one to translate semiclassical dynamics
into a closed system of partial differential equations (PDEs) \cite{CGHS}.
Although the exact solution to these PDEs is not known explicitly,
it is possible to explore these solutions numerically, and also through
certain analytical approximations (see below) \cite{Aprox}.

The purpose of this paper is to explore \emph{static} BH solutions
\cite{static} of the 2D semiclassical CGHS model,
\footnote{From the physical view-point, a static semiclassical BH solution should
be viewed as the (somewhat hypothetical) situation in which a constant
(quantum) influx arrives from past null infinity and falls into a pre-existing BH,
exactly compensating the Hawking outflux. It is a generalization of the
Hartle-Hawking quantum state to the CGHS framework.}
and to find out how the surface gravity (and hence also temperature
and outflux) changes with the BH mass $M$, due to the semiclassical
backreation on the metric, in the domain of large mass ($M\gg m_{p}$).
We shall address this problem here both numerically and analytically.
Note that in the static case the CGHS field equations reduce to ordinary
differential equations (ODEs), which drastically simplifies their
numerical (as well as approximate-analytical) solution.
Also, to address our problem it will be sufficient to analyze the exterior part of the BH.
At the analytical
side, since for large M the semiclassical BH-exterior geometry is well-approximated
(locally) by the corresponding classical solution, we shall consider
the semiclassical solution as a small deviation from the classical
one, and treat this deviation by linear perturbation analysis. As
it turns out, the overall magnitude of this perturbation scales as
$1/M$. Not surprisingly, the perturbation analysis yields a semiclassical
correction to $\kappa$ (as well as to $T_{H}$ and the outflux) which
scales as $1/M$ too. We calculate this correction analytically, and
then confirm it numerically.

Note that there is a remarkable difference between 2D and 4D BHs,
already at the classical level: Whereas in 4D the surface gravity
$\kappa$ scales as $1/M$, in 2D it is actually \emph{independent
of the BH mass}. As a consequence, in a semiclassical macroscopic
BH the (leading-order) Hawking temperature $T_{H}$ is $\propto1/M$
in 4D, but constant (i.e. independent of $M$) in the 2D framework.
Correspondingly the outflux (the outgoing component of $\hat{T}_{\alpha\beta}$)
at infinity is $\propto M^{-2}$ in 4D but constant in 2D. The semiclassical
correction to the background geometry of the 2D BH modifies these
constant values of $\kappa$, $T_{H}$ and outflux, and the modification
in all three quantities scales as $1/M$ at the leading order. It
is this $\propto1/M$ leading-order semiclassical effect which we
explore in this paper---both theoretically (through perturbation analysis)
and numerically.

We point out that a similar phenomenon also occurs in the case of
a \emph{dynamical, evaporating}, 2D CGHS BH. At the large-mass limit,
a 2D BH evaporates at a constant rate $\dot{M}$ (owing to the $M$-independence
of $\kappa$ in the classical CGHS solution). However, due to the
backreaction of the semiclassical $\hat{T}_{\alpha\beta}$ on the
geometry, there is a finite-mass correction to the Hawking outflux
(and hence to $\dot{M}$), which again scales as $1/M$.
\footnote{In the evaporating case, this ``$M$'' actually refers to the
Bondi mass, namely the remaining BH mass as seen by an observer at
FNI.
} This correction term for a 2D evaporating BH was calculated analytically
\cite{Aprox1}, and also confirmed numerically \cite{APR,Dori_Ori}.
It is remarkable that the leading-order ($\propto1/M$) finite-mass
correction to the (otherwise-constant) outgoing component of $\hat{T}_{\alpha\beta}$
is found to be \emph{exactly the same} in the static and evaporating
cases. We further comment on this observation in the Discussion section.

In the next section we briefly review the CGHS semiclassical model
(as well as its classical limit). We restrict attention to static
solutions which are regular at the horizon, and explore the asymptotic
behavior of the semiclassical geometry at the horizon and at infinity.
We then turn in section 3 to address the Hawking temperature of such
static BH's, as well as the Hawking outflux (and influx) at infinity,
taking into account the deviation of the background geometry from
its classical counterpart. Then in Sec. 4 we analyze the large-mass
leading-order ($\propto1/M$) semiclassical correction to the geometry,
treating it as a linear perturbation. This in turn yields the $\propto1/M$
correction to $\kappa$, and hence to the BH temperature $T_{H}$
and the Hawking outflux. These analytical results are verified numerically
in Sec. 5. Finally in Sec. 6 we briefly discuss our results, and compare
them to the corresponding case of 2D evaporating BH.

\section{The CGHS model}

The CGHS model \cite{CGHS} consists of a two-dimensional metric $g_{\alpha\beta}$
coupled to a dilaton $\phi$ and to a large number $N\gg1$ of identical minimally-coupled,
massless, scalar fields $f_{i}$. It is convenient to express the metric
in the conformal form, \begin{equation}
g_{uv}=-\frac{1}{2}e^{2\rho},\,\; g_{uu}=g_{vv}=0.\label{eq:Metric}\end{equation}
At the classical level, the action then takes the form \begin{equation}
S_{classic}=\frac{1}{\pi}\int dudv\left[e^{-2\phi}\left(-2\rho_{,uv}+4\phi_{,u}\phi_{,v}-\lambda^{2}e^{2\rho}\right)-\frac{1}{2}\sum_{i=1}^{N}f_{i,u}f_{i,v}\right]\label{eq:Classic_Action}\end{equation}
where $\lambda$ is the cosmological constant, and we set $\lambda=1$ henceforth.
\footnote{We can always absorb the factor $\lambda ^2$ in the action by a change of variable $\rho\to\tilde{\rho}=\rho+\ln{\lambda}$, which does not affect the field equations. This may actually be associate to a choice of length unit. Note that our choice $c=G=\lambda=1$ completely fixes the unit system.}
At the semiclassical level, the trace
anomaly contributes an effective term \cite{CGHS} \[
S_{trace}=\frac{N}{12\pi}\int\rho_{,u}\rho_{,v}dudv,\]
leading to the overall semiclassical action \begin{equation}
S_{sc}=S_{classic}+S_{trace}.\label{eq:S-C_Action}\end{equation}

Variation of $S_{sc}$ with respect to the scalar fields $f_{i}$
yields the standard wave equation $f_{i,uv}=0$, and throughout the
paper we shall consider the trivial solution $f_{i}=0$. Variation
of $\phi$ yields a certain nonlinear hyperbolic equation, and the
variation of the metric yields one hyperbolic equation and two additional
constraint equations. Overall, there are two evolution equations and
two constraint equations (which will be presented shortly).

The coordinate transformations which preserve the double-null structure
(\ref{eq:Metric}) of the line element are of the form $u\to\tilde{u}(u),v\to\tilde{v}(v)$.
They transform $\rho$ into \[
\tilde{\rho}=\rho-\frac{1}{2}\ln\frac{d\tilde{u}}{du}-\frac{1}{2}\ln\frac{d\tilde{v}}{dv}.\]

Before analyzing the semiclassical dynamics, it will be useful to
consider the (much simpler) classical system, construct its general
static solution, and explore its asymptotic properties. The insights
gained from the classical system will in turn facilitate our analysis
of the more complex semiclassical dynamics.

\subsection{Classical equations}

The classical field equations are obtained by varying $S_{classic}$
with respect to the dilaton and the metric. These equations are much
simplified by the fields redefinition \[
R\equiv e^{-2\phi},\;\; S\equiv2(\rho-\phi).\]
The evolution equations then take the form \begin{equation}
R_{,uv}=-e^{S},\,\; S_{,uv}=0.\label{eq:Classic_evo_eq,2d}\end{equation}
In addition there are two constraint equations, which (after substituting
$f_{i}=0$) read \begin{equation}
R_{,u}S_{,u}-R_{,uu}=0\,,\;\; R_{,v}S_{,v}-R_{,vv}=0  .
\label{eq:Classic_const_eq_2d}\end{equation}

Note that \[
\rho=(S-\ln R)/2.\]
In a coordinate transformation $u\to\tilde{u}(u),v\to\tilde{v}(v)$,
$R$ is unchanged but $S$ transforms according to \begin{equation}
\tilde{S}=S-\ln\frac{d\tilde{u}}{du}-\ln\frac{d\tilde{v}}{dv}.\label{eq:S-transform}\end{equation}
The form of the evolution and constraint equations is unchanged by
such a transformation.

The general solution of this system of classical field equations (evolution+constraints)
is known to be uniquely described by a one-parameter family of solutions
(up to coordinate transformation), which in the so-called \emph{Eddington-like
coordinates} take the form \begin{equation}
R=M+e^{v-u},\;\; S=v-u,\label{eq:VacuumClassic_uv}\end{equation}
where $M$ is an arbitrary constant. For $M>0$ (which we shall assume
throughout) these solutions describe a black hole (BH) with ADM mass
$M$.

Throughout this paper we shall restrict attention to \emph{static}
(classical and semiclassical) solutions, namely, solutions which only
depend on the spatial variable $x\equiv v-u$. All field equations
then reduce to ordinary differential equations (ODEs), which drastically
simplifies the analysis. The classical evolution equations then read
\begin{equation}
R''=e^{S},\;\; S''=0,\label{eq:Classic_evo_eq_1d}\end{equation}
and the two constraint equations reduce to a single one: \begin{equation}
R'S'-R''=0,\label{eq:Classic_const_eq_1d}\end{equation}
where a prime denotes differentiation with respect to x. The general
classical solution (\ref{eq:VacuumClassic_uv}) is manifestly static, and we re-write it as \begin{equation}
R=e^{x}+M,\;\; S=x.\label{eq:VacuumClassic_x}\end{equation}

In this classical vacuum solution \[
\rho=\frac{1}{2}[x-\ln(e^{x}+M)].\]
Note that $\rho$ vanishes at the limit $x\to\infty$ (corresponding
to spacelike as well as null infinity), implying that the classical
BH spacetime is asymptotically flat.

The other asymptotic limit, $x\to-\infty$, corresponds to
approaching the black-hole horizon. At that limit $R$ decays exponentially
to $\mbox{M}$, whereas $S\to-\infty$.
\footnote{The divergence of $x$ and $S$ (and $\rho$) at the horizon merely
represents a coordinate singularity. To remove this singularity one
may transform from the Eddington-like coordinates $u,v$ to new, Kruskal-like
coordinates $U=-e^{-u}$ and $V=e^{v}$. \label{fn:Kruskal}
}

To facilitate the semiclassical analysis below, it will be useful
to explore the general (static) solution of the classical evolution system
(\ref{eq:Classic_evo_eq_1d}), while relaxing the constraint equation
(\ref{eq:Classic_const_eq_1d}). We shall refer to this more general
solution of Eqs. (\ref{eq:Classic_evo_eq_1d}) as the (static) classical
\emph{flux-carrying }solution --- to be distinguished from the strict
\emph{vacuum }solution (\ref{eq:VacuumClassic_x}) (which satisfies
the constraint equations as well).
\footnote{More generally, a classical (non-static) flux-carrying\emph{ }solution
will be defined to be a solution of the evolution equations (\ref{eq:Classic_evo_eq,2d})
which does not necessarily satisfy the constraint equations (\ref{eq:Classic_const_eq_2d}). }
As we shall see below, in the two important asymptotic regions---horizon
and infinity---the semiclassical solutions are well approximated by
certain flux-carrying classical solutions (which mimic the semiclassical
fluxes in these two asymptotic regions).

The integration of the evolution equations (\ref{eq:Classic_evo_eq_1d})
is straightforward, and one obtains the general solution \begin{equation}
R=a^{-2}e^{ax+b}+\tilde{c}x+\tilde{m},\,\; S=ax+b,\label{eq:Classic_Static_sol,ungauged}\end{equation}
with four arbitrary constants $a,b,\tilde{c},\tilde{m}$ (as expected
for two second-order ODEs). However, by a simple, linear coordinate
transformation of the form $u\to u'(u)$, $v\to v'(v)$
one can get rid of two parameters, say $a$ and $b$,
\footnote{The coordinate transformations which preserve the double-null form
of the metric, as well as staticity (namly, dependence on $\tilde{x}=\tilde{v}-\tilde{u}$
only) must be of the linear form
$\tilde{u}=\alpha u+\beta_{u},\tilde{v}=\alpha v+\beta_{v}$
(such that $ $$\tilde{x}=\alpha x+\beta$, where $\beta=\beta_{v}-\beta_{u}$).
By appropriate choice of $\alpha$ and $\beta$ one can get rid of
$a$ and $b$ in Eq. (\ref{eq:Classic_Static_sol,ungauged}), which
also modifies $\tilde{c}$ and $\tilde{m}$. [Note that in this
coordinate transformation $R$ is unchanged and $S$ changes into
$\tilde{S}=S-2\ln\alpha$; However, to address the transition from
(\ref{eq:Classic_Static_sol,ungauged}) to (\ref{eq:Classic_Static_sol,gauged})
one needs to consider the \emph{functional form} of $R(\tilde{x})$
and $\tilde{S}(\tilde{x})$.] \label{fn:transformation}}
bringing the solution to its canonical form: \begin{equation}
R=e^{x}+cx+m,\;\; S=x.\label{eq:Classic_Static_sol,gauged}\end{equation}
Thus, the static classical flux-carrying solution is actually a two-parameter
family, in which $c$ represents the flux magnitude (as we shortly
discuss), and the other parameter $m$ is reminiscent of the mass
parameter [compare to Eq. (\ref{eq:VacuumClassic_x})].

For any configuration $R(x),S(x)$ we define the \emph{flux} $T(x)$
to be the differential expression at the left-hand side of the constraint
equation (\ref{eq:Classic_const_eq_1d}), namely $T\equiv R'S'-R''$.
In a flux-carrying solution, the evolution equation (\ref{eq:Classic_evo_eq_1d})
guarantees that $T=const$, as one can easily verify
\footnote{More generally, for any field configuration $R(u,v),S(u,v)$ one may
define the fluxes $T_{uu}$ and $T_{vv}$ to be the corresponding
two differential expressions in the constraint equations (\ref{eq:Classic_const_eq_2d}).
If the evolution equation (\ref{eq:Classic_evo_eq,2d}) is imposed,
it implies that $\partial_{v}T_{uu}=\partial_{u}T_{vv}=0$. Then,
if the solution is further assumed to be static, $T_{uu}=T_{vv}=T$
and the latter must be a constant.)
}. Expressing the flux-carrying solution in its canonical form (\ref{eq:Classic_Static_sol,gauged}),
one readily finds that $T=c$.

The spacetime's geometry is fully dictated by $\rho(x)$, which may
be expressed as \begin{equation}
\rho=-\frac{1}{2}\ln\left[1+(cx+m)e^{-x}\right].\label{eq:RhoClassical}\end{equation}
At spacelike infinity ($x\to\infty$), $\rho\propto(cx+m)e^{-x}$.
In particular this implies that \begin{equation}
\rho,\rho',\rho''\to0,\qquad\qquad(x\to+\infty)\label{eq:ClassicInfinity}\end{equation}
with all three quantities at the left-hand side decaying exponentially
in $x$. Thus, as long as the strict spacetime metric is concerned,
the classical solutions (vaccum as well as flux-carrying) are all
asymptotically-flat: the metric approaches 2D Minkowski ($\rho=0$),
and the curvature asymptotically vanishes.

Next we consider the asymptotic behavior of the flux-carrying solutions
at the horizon limit ($x\to-\infty$), starting with the $c<0$ case. It will be instructive
to deduce this asymptotic behavior directly from the evolution equations
(\ref{eq:Classic_evo_eq_1d}) (rather than by just substituting $x\to-\infty$
in the explicit solution), because the same procedure will later serve
us in the semiclassical case, wherein we shall analyze the near-horizon asymptotic
behavior of the semiclassical evolution equations (see below).

The equation $S''=0$ dictates $S'=const$, yielding $S\to-\infty$
at the horizon. This, in turn, implies that the source term $e^{S}$
at the right-hand side of the evolution equation for $R$ dies out,
leading to $R''\cong0$ and hence to a linear asymptotic behavior
of $R(x)$ at $ $the horizon. This is of course confirmed by the
linear asymptotic form of the explicit solution (\ref{eq:Classic_Static_sol,gauged})
as $x\to-\infty$.

The Ricci scalar $R_{\alpha}^{\alpha}$ is given by $-4e^{-2\rho}\rho''$.
At the limit $x\to-\infty$ one finds
\begin{equation}
R_{\alpha}^{\alpha}\cong-\frac{2c}{x}e^{-x}\to\infty.\label{eq:RicciClassical}\end{equation}
[We also note that the scalar $(\nabla\phi)^{2}$ too diverges like
$\propto(c/x)e^{-x}$ as $x\to-\infty$.] Thus, all
flux-carrying solutions with $c<0$ develop a curvature singularity
at the  ``horizon'' limit $x\to-\infty$.

The solutions with $c>0$ are singular too. In this case, however, the singularity occurs at {\it finite} negative $x$ (rather than $x\to-\infty$), at the point where $R$ vanishes [cf. Eq. (\ref{eq:Classic_Static_sol,gauged})] and $\rho$ diverges.
Thus, the only regular (static) BH solutions are those with $c=0$ (and $M>0)$, namely the
true vacuum BH solutions. The $c=0$ solutions are perfectly regular
at the horizon, and can be extended to the interior domain $R<M$
in the usual procedure, by transformation to Kruskal-like coordinates
(see footnote \ref{fn:Kruskal}).

For later convenience we summarize here the (classical) $x\to-\infty$
asymptotic behavior as follows: There is a one-parameter regular class
which behaves as \begin{equation}
R\cong m\;,\;\; R'=O(e^{x})\;,\;\; S=x;\label{eq:ClassicalHorizonRegular}\end{equation}
And there also exists a larger, two-parameter class, which behaves
as \begin{equation}
R\cong cx+const\;,\;\; R'\cong c\;,\;\; S=x,
\label{eq:ClassicalHorizonSingular}\end{equation}
which however admits a curvature singularity rather than a regular horizon.

\subsection{Semiclassical equations}

At the semiclassical level the field equations get quantum correction
terms coming from $S_{trace}$. The semiclassical evolution equations
become

\begin{equation}
R_{,uv}=-e^{S}-K\rho_{,uv},\;\; S_{,uv}=K\rho_{,uv}/R,\label{eq:Semi-Classic_evo_eq,2d}\end{equation}
where $K\equiv N/12$. The two constraint equations now take the form

\begin{equation}
R_{,w}S_{,w}-R_{,ww}=K\left[\rho_{,ww}-\rho_{,w}^{2}+z_{w}\left(w\right)\right]\equiv\hat{T}_{ww}\,,\label{eq:Semi-Classic_const_eq,2d}\end{equation}
where hereafter $w$ will stand for either $u$ or $v$, and by $\hat{T}_{ww}$
we refer to the $ww$ component of the renormalized stress-energy
tensor. The functions $z_{u}(u)$,$z_{v}(v)$ carry the information
about the quantum state, and should thus be determined by initial
or boundary conditions.

In the static model, where both $R$ and $S$ depend on $x=v-u$ only,
the evolution equations reduce to the ODEs

\begin{equation}
R''=e^{S}-K\rho'',\,\; S''=K\rho''/R.\label{eq:Semi-Classic_evo_eq,1d}\end{equation}
The constraint equations again reduce to a single ODE:

\begin{equation}
R'S'-R''=K\left[\rho''-\rho'^{2}+z\right]\equiv\hat{T},\label{eq:Semi-Classic_const_eq,1d}\end{equation}
where now $\hat{T}=\hat{T}_{vv}=\hat{T}_{uu}$. Note that in the static
case the $w=u$ version of Eq. (\ref{eq:Semi-Classic_const_eq,2d})
implies $z=z_{u}(u)$, and similarly its $w=v$ version implies $z=z_{v}(v)$,
hence in Eq. (\ref{eq:Semi-Classic_const_eq,1d}) $z$ must be a constant
(it will be determined below by the regularity condition at the horizon).

\subsubsection{Far-field asymptotic behavior}

The asymptotic behavior of the semiclassical solution at infinity
turns out to be fairly similar to the (flux-carrying) classical solution.
Although we cannot prove this rigorously, this conclusion is easily
derived from a simple iterative procedure, starting from the classical
solution. Since in the latter $\rho''$ decays exponentially in $x$,
we can drop $\rho''$ in the evolution system (\ref{eq:Semi-Classic_evo_eq,1d}).
The latter then reduces to the classical evolution system (\ref{eq:Classic_evo_eq_1d}),
which we have already integrated in the previous subsection. We find
that the far-field semiclassical asymptotic behavior again takes the
general form

\begin{equation}
R\cong e^{x}+\hat{c}x+\hat{m},\;\;
S\cong x\qquad\qquad(x\to+\infty)\label{eq:S-C_RS_inf_asymptotic}
\end{equation}
(with exponentially-small corrections), where $\hat{c}$ and $\hat{m}$
are free parameters
\footnote{Just like in the classical case, the general solution of the evolution
equations admits four free parameters, but by applying a gauge transformation
we get rid of two parameters (see footnote \ref{fn:transformation}
above) and obtain the canonical form (\ref{eq:S-C_RS_inf_asymptotic}).
\label{fn:canonical}}.
In particular, in the semiclassical case $\rho(x)$ again satisfies
the asymptotic behavior \begin{equation}
\rho,\rho',\rho''\to0,\qquad\qquad(x\to+\infty)\label{eq:SemiclassicInfinity}\end{equation}
with all three quantities decaying exponentially (which in retrospect
guarantees the consistency of this far-field approximation).

Note that when the asymptotic expression (\ref{eq:S-C_RS_inf_asymptotic})
is substituted in the semiclassical constraint equation (\ref{eq:Semi-Classic_const_eq,1d}),
it yields
\footnote{By iteration one finds that the large-$x$ corrections to Eq. (\ref{eq:S-C_RS_inf_asymptotic}) are
$O(x e^{-x} )$ for $R$ and $O(x e^{-2x}) $ for $S$
(with a pre-factor $ \propto K \hat{c}$ in both cases),
which at $x\to\infty$ yield no contribution to
Eq. (\ref{eq:Semi-Classic_const_eq,1d}).} \[
z=\hat{c}/K.\]

\subsubsection{Near-horizon asymptotic behavior}

We turn now to analyze the semiclassical asymptotic behavior at the
horizon limit $x\to-\infty$. To this end, it is useful to
represent the evolution equations in their ``canonical'' form
(where $R''$ and $S''$ are explicitly expressed in terms of $R$
and $S$). Substituting $\rho=(S-\ln R)/2$ in Eq. (\ref{eq:Semi-Classic_evo_eq,2d})
and then extracting $R''$ and $S''$, one obtains

\begin{equation}
R''=e^{S}\frac{2R-K}{2\left(R-K\right)}-R'^{2}\frac{K}{2R\left(R-K\right)}\label{eq:S-C_Canonic_evo_R}\end{equation}
and\begin{equation}
S''=-e^{S}\frac{K}{2R\left(R-K\right)}+R'^{2}\frac{K}{2R^{2}\left(R-K\right)}.\label{eq:S-C_Canonic_evo_S}\end{equation}
To analyze these equations we again apply an iterative procedure, starting
from the (regular) classical near-horizon asymptotic behavior (\ref{eq:ClassicalHorizonRegular}).
Since both $R'$ and $e^{S}$ decay exponentially on approaching the
horizon ($x\to-\infty$), in the two evolution equations (\ref{eq:S-C_Canonic_evo_R},\ref{eq:S-C_Canonic_evo_S})
the right-hand side decays like $e^{x}$ too, namely $R''=S''=O(e^{x})$.
Integrating these equations, one finds that both $R'$ and $S'$ approach
finite values at the horizon:
$R'\to const\equiv R_{1},S'\to const\equiv S_{1}$,
where $R_{1}$ and $S_{1}$ are yet arbitrary.
However, one again finds that as long as $R_{1}\neq0$, a singularity develops (at a finite proper distance) instead of a regular horizon:
For $R_{1}<0$, $R$ diverges to $+\infty$ as $x\to-\infty$, and the Ricci scalar $R_{\alpha}^{\alpha}$
diverges like $\propto R_{1}e^{-x}\to\infty$ at that limit
[in full analogy with the classical case, Eq. (\ref{eq:RicciClassical}) above]. For $R_{1}>0$, $R$ decreases monotonically until it eventually approaches $K$, yielding again a curvature singularity (closely related to the type analyzed in Ref. \cite{Levanony}).
For obvious physical reasons we restrict our attention here to \emph{horizon-regular}
semiclassical BH solutions, and therefore we set $R_{1}=0$. Doing
so, we obtain the general regular near-horizon asymptotic behavior:
\begin{equation}
R\cong const\equiv\bar{m}\;,\;\; S\cong S_{0}+S_{1}x,\label{eq:Near-Horizon_RS}\end{equation}
with $O(e^{x})$ deviations. The derivatives of $R$ and $S$ satisfy
\begin{equation}
S'\to const\equiv S_{1}\label{eq:Near-Horizon_dS}\end{equation}
and \begin{equation}
R'=R''=S''=O(e^{x}).\label{eq:Near-Horizon_ddRS}\end{equation}
Setting $\rho'=(S'-R'/R)/2$, one finds at the horizon limit:

\begin{equation}
\rho'\to const=\frac{S_{1}}{2}\;,\;\;\rho''\to0.\qquad\qquad(x\to-\infty)\label{eq:RhoHorizon}\end{equation}
Recall also that in the classical solution $S_{1}=1$.

It is not difficult to see that this asymptotic behavior (with $S_{1}>0$; see below), along with the asymptotic-flatness condition (\ref{eq:SemiclassicInfinity}), indeed guarantees the presence of a black hole, with its horizon located at $x\to -\infty$ (more specifically, future event horizon at $u\to\infty$ and past horizon at $v\to -\infty$).
\footnote{To this end one should bear in mind that no singularity of $R$ or $S$ occurs at finite $x$ in the domain considered here (assuming $\bar m>K$)---a fact which we verified numerically but can also be shown analytically. This in particular implies that neither a horizon nor a spacetime singularity can develop at any finite $x$.}

A few remarks are in order here: (i) In principle we could easily
bring the value of $S_{1}$ to unity by a coordinate transformation,
as we have done in the classical solution. Such a transformation,
however, would multiply $S'(x)$ by a global parameter (the constant
$1/\alpha$ in the terminology of footnote \ref{fn:transformation}).
But we have already fixed this constant by demanding $S'=1$
at spacelike infinity, as demonstrated in Eq. (\ref{eq:S-C_RS_inf_asymptotic})
(see also footnote \ref{fn:canonical});
hence in the semiclassical case we no longer have the freedom to gauge
out the near-horizon parameter $S_{1}$. (ii) $\bar{m}$ is a natural
extension of the classical mass parameter $m$ to the semiclassical
case. (iii) The above near-horizon analysis is restricted to the case
$\bar{m}>K$: Then, when $R$ approaches $\bar{m}$ at the horizon,
the terms $R-K$ in the denominators in Eqs. (\ref{eq:S-C_Canonic_evo_R},\ref{eq:S-C_Canonic_evo_S})
remain bounded away from zero. (iv) In fact, throughout the rest of this paper
we are primarily interested in the macroscopic case, namely $\bar{m}\gg K.$
In this case, the solution is well approximated by the classical solution.
In the latter $S_{1}\equiv S'=1$, as seen in e.g. Eq. (\ref{eq:Classic_Static_sol,gauged}).
For this reason, in a macroscopic BH one expects that $S_{1}\approx1$
(this will be shown more explicitly in the next section, in which
the deviation of $S_{1}$ from unity is explicitly calculated in the
large-mass case and shown to vanish at the macroscopic limit). In
particular, it follow that $S_{1}$ is \emph{positive} (at least in the large-mass
case).

We also point out that the asymptotic behavior (\ref{eq:Near-Horizon_dS},\ref{eq:Near-Horizon_ddRS})
guarantees that $R'$ and $e^{S}$ indeed decay exponentially on approaching
the horizon, which in retrospect justifies our near-horizon approximation
[in which we have regarded the right-hand sides of Eqs. (\ref{eq:S-C_Canonic_evo_R},\ref{eq:S-C_Canonic_evo_S}) as negligible].

\subsubsection{Concluding remark}

In the previous section we asserted that the flux-carrying
classical solution well approximates the semiclassical solution
at both asymptotic boundaries, namely at the far-field region and near the horizon.
We are now in a position
to verify and further clarify this statement, by comparing the exact
classical solution (\ref{eq:Classic_Static_sol,gauged}) to the semiclassical
asymptotic solutions at $x\to\infty$ and $x\to-\infty$.

In the far-field region, Eq. (\ref{eq:S-C_RS_inf_asymptotic}) conforms
with (\ref{eq:Classic_Static_sol,gauged}), with the mapping $c\to\hat{c},m\to\hat{m}$
(and, as will become clear in the next section, $\hat{c}\neq0$ in
the semiclassical case---which is the essence of the Hawking effect).
Near the horizon, Eq. (\ref{eq:Near-Horizon_RS}) conforms with Eq.
(\ref{eq:Classic_Static_sol,gauged}), but this time with the mapping
$c\to0,m\to\bar{m}$.

So here is a key difference between the classical and semiclassical
solutions: In the classical case, the exact flux-carrying solution
(\ref{eq:Classic_Static_sol,gauged}) holds all the way from infinity
to the horizon, with a \emph{single} set of parameters $c,m$. Thus,
regularity at the horizon ($c=0$) automatically guarantees vanishing
flux at infinity (and everywhere).
On the contrary, in the semiclassical case,
although Eq. (\ref{eq:Classic_Static_sol,gauged}) is a good
approximation both at the near-horizon and far-field regions, it fails
to approximate the solution in between, and the consequence is that
the free parameters $c,m$ take different values at the two asymptotic
regions. In particular, the difference in $c$ (namely $c=0$ at the
horizon but $c\neq0$ in the far-field region) is a manifestation
of the Hawking effect, as we further discuss in the next section.

\emph{Notational remark}: Throughout the rest of this paper, we shall
replace the symbol $\bar{m}$ by $m$ for notational simplicity. Thus,
by the mass $m$ of a semiclassical BH, we shall specifically refer
to the value of $R$ at the horizon.

\section{Hawking temperature and the semiclassical fluxes }

\subsection{Surface gravity and Hawking temperature}

The Hawking temperature $T_{H}$ of a static BH is directly dictated
by the horizon's \emph{surface-gravity} $\kappa$. When the 2D metric
is expressed in the Schwarzschild-like form
$$ds^{2}=-F(r)dt^{2}+[1/F(r)]dr^{2},$$
the surface gravity is given by the standard expression
\begin{equation}
\kappa=\frac{1}{2}\left(\frac{dF}{dr}\right)_{(horizon)}.\label{eq:kappa_standard}\end{equation}
In the double-null metric we use here, $\kappa$ takes the simple form

\begin{equation}
\kappa=2\rho'_{(horizon)}.\label{eq:kappa_general}\end{equation}
[To verify this, one may start from the above Schwarzschild-like
line element, define $r_{*}(r)$ through $dr/dr_{*}=F(r),$ and then
introduce the null coordinates $v=t+r_{*},u=t-r_{*},$ recovering the
metric (\ref{eq:Metric}) with $e^{2\rho}=F(r)$. Noting that $x=2r_{*}$,
one obtains Eq. (\ref{eq:kappa_general}).]

In our 2D semiclassical model, setting $2\rho'=S'-R'/R$ and recalling
the near-horizon asymptotic behavior (\ref{eq:Near-Horizon_dS},\ref{eq:Near-Horizon_ddRS}),
we obtain \begin{equation}
\kappa=S_{1}.\label{eq:kappa_S1}\end{equation}

Note that $\rho'$ in Eq. (\ref{eq:kappa_general}) is in fact gauge-dependent,
and the right-hand side of Eq. (\ref{eq:kappa_general}) is to be
evaluated using the ``asymptotically-Minkowskian gauge'', namely
the specific gauge in which $\rho$ vanishes at infinity [which is
indeed the gauge we use here, cf. Eq. (\ref{eq:SemiclassicInfinity})].
For later convenience we also express $\kappa$ in a gauge-invariant
manner: \begin{equation}
\kappa=\frac{2\rho'_{(-\infty)}}{S'_{(\infty)}}=\frac{S'_{(-\infty)}}{S'_{(\infty)}},\label{eq:kappa_general-1}\end{equation}
where $S'_{(\pm\infty)}\equiv S'(x\to\pm\infty)$ etc.: In
a gauge transformation $\tilde{x}=\alpha x+\beta$ (cf. footnote \ref{fn:transformation}),
the difference $ds$ (or $d\rho$) associated with two adjacent spacetime points is
unchanged, whereas $dx$ changes by the factor $\alpha$, which however
cancels out between the numerator and denominator.

Once $\kappa$ is obtained, the Hawking temperature $T_{H}$ is given
by \cite{Zaslavskii}
\begin{equation}
T_{H}=\kappa/2\pi.\label{eq:temp-1}\end{equation}
This relation may be derived straightforwardly by analytically extending
the 2D metric (e.g. in the above Schwarzschild-like form) into the
euclidean domain, just like Hawking's original derivation \cite{Hawking}
in the 4D case.

At the classical limit $\kappa=S_{1}=1$. It is remarkable that unlike
the 4D Schwarzschild BH, wherein $\kappa\propto1/m$, in the classical
2D case $\kappa$ is \emph{independent of }$m$ --- and so is the
Hawking temperature.

\subsection{Semiclassical fluxes at infinity}

As was discussed above, in a static solution the semiclassical outflux
and influx are equal at each point: $\hat{T}_{vv}=\hat{T}_{uu}$$\equiv\hat{T}(x)$.
One of our main goals in this paper is to calculate the value of $\hat{T}$
at infinity, which we shall denote $\hat{T}_{\infty}$.

Equations (\ref{eq:Semi-Classic_const_eq,1d}) and (\ref{eq:SemiclassicInfinity})
yield $\hat{T}_{\infty}=Kz$. The parameter $z,$ in turn, may be
determined from the asymptotic behavior at the horizon: Recalling
Eqs. (\ref{eq:Near-Horizon_dS},\ref{eq:Near-Horizon_ddRS}), the
left-hand side of Eq. (\ref{eq:Semi-Classic_const_eq,1d}) vanishes,
which by virtue of Eqs. (\ref{eq:RhoHorizon}) and (\ref{eq:kappa_S1})
implies \[
z=\frac{S_{1}^{2}}{4}=\frac{\kappa^{2}}{4}.\]
The fluxes at infinity are therefore \begin{equation}
\hat{T}_{\infty}=K\frac{S_{1}^{2}}{4}=K\frac{\kappa^{2}}{4}.\label{eq:Flux_Exact}\end{equation}

Alternatively the outflux $\hat{T}_{\infty}$ at asymptotic infinity
may be obtained from the Hawking temperature $T_{H}$. Owing to asymptotic
flatness of the 2D metric [cf. Eq. (\ref{eq:SemiclassicInfinity})],
we may apply at asymptotic infinity the standard rules of flat-space
2D physics
\footnote{Note that in the CGHS 2D model the scalar fields $f_{i}$ do not admit
any scattering due to curvature, hence there is no ``gray factor''
in the Hawking outflux (unlike the 4D case). }
: For each of the $N$ scalar fields $f_{i}$, the outflux is related
to the temperature through the 2D analog of the Stefan-Boltzmann
law: $\hat{T}_{\infty}^{i}=(\pi^{2}/12)T_{H}^{2}$ . Multiplying by
$N=12K$ and recalling Eq. (\ref{eq:temp-1}), one recovers Eq. (\ref{eq:Flux_Exact}).

Thus, all that is needed for the determination of $\hat{T}_{\infty}$
is $S_{1}$, namely the horizon value of $S'$. At the macroscopic
limit ($m\to\infty$), one can employ the classical solution
for this purpose, yielding $S_{1}=1$ and hence $\hat{T}_{\infty}=K/4$.
For a finite $m$ there will be a deviation from this classical value
of $S_{1}$, which will grow with decreasing mass. In the next section
we shall calculate this deviation for large $m$, at the first order
in $m/K$.

\section{Finite-mass semiclassical corrections}

The spacetime outside of a macroscopic semiclassical BH is (in a local
sense) well approximated by the classical solution. This is most easily
seen in the evolution equation (\ref{eq:S-C_Canonic_evo_S}) for $S$,
wherein the semiclassical correction is proportional to $K.$ When
substituting the classical expressions for $R$ and $S$ in the right-hand
side, and taking the leading order in $K/R$, one recognizes that each of
the two terms in the right-hand side is bounded by $K/2R$---which
in turn is bounded, outside the BH, by $\sim K/m$. We may therefore
expect that the semiclassical correction to $S$ should be small whenever
$m\gg K$. Inspecting the evolution equation (\ref{eq:S-C_Canonic_evo_R})
for $R$ leads to a similar conclusion: The semiclassical contribution
to $R''$ is smaller than the classical contribution ($e^{S}$) by
a small factor of order $K/R\lesssim K/m$.
This conclusion---namely the smallness of the semiclassical correction to both $R$ and
$S$ in the case $m\gg K$---was also verified in our numerical simulations
(described in the next section).

Thus, to investigate the small semiclassical correction in the case
of a large-mass BH, it is useful to express the solution in the form
\[
R(x)=R_{cl}(x)+\delta R(x)\;,\;\; S(x)=S_{cl}(x)+\delta S(x),\]
where $R_{cl}$ and $S_{cl}$ denote the classical solution (\ref{eq:VacuumClassic_x}).
The semiclassical corrections $\delta R$ and $\delta S$ can then
be treated as small perturbations. Linearizing the evolution equations
(\ref{eq:S-C_Canonic_evo_R},\ref{eq:S-C_Canonic_evo_S}) in $\delta R$
and $\delta S$ (and in $K$), one obtains the inhomogeneous linear
system \begin{equation}
\delta R''=\delta S\, e^{S_{cl}}+\frac{K}{2R_{cl}}\left[e^{S_{cl}}-\frac{(R'_{cl})^{2}}{R_{cl}}\right]\label{eq:deltaR''}\end{equation}
and \begin{equation}
\delta S''=\frac{K}{2R_{cl}^{3}}\left[(R'_{cl})^{2}-R_{cl}e^{S_{cl}}\right].\label{eq:deltaS''}\end{equation}
Substituing the explicit expressions (\ref{eq:VacuumClassic_x})
for $R_{cl}$ and $S_{cl}$, the system takes the form
\begin{equation}
\delta R'' = \delta S{e^x} + K{{m{e^x}} \over {2{{\left( {{e^x} + m} \right)}^2}}}
\label{eq:deltaR''_Explicit}
\end{equation}

and
\begin{equation}
\delta S''=-\frac{Km}{2}\frac{e^{x}}{\left(e^{x}+m\right)^{3}}.\label{eq:deltaS''_Explicit}
\end{equation}

The integration of this system is straightforward, and the full explicit solution is presented in the Appendix.
Here we shall only need the equation for $\delta S$, because
the parameter $\kappa=S_{1}$ (which determines the Hawking temperature
and outflux) is entirely determined by $\delta S'(x)$. It is convenient that Eq.
(\ref{eq:deltaS''}) for $\delta S''$ is decoupled from $\delta R$. Its first integral yields
\begin{equation}
\delta S'=\frac{Km}{4\left(e^{x}+m\right)^{2}}.\label{eq:delta_S',fina_1st order}\end{equation}
Notice that we have chosen the integration constant such that $\delta S'$
vanishes at $x\to\infty$, to comply with the canonical form
(\ref{eq:S-C_RS_inf_asymptotic}) of the asymptotic behavior at spacelike
infinity.

At the horizon limit we obtain
\[S_{1}=1+\delta S'(x\to-\infty)=1+K/(4m).\]
Thus, up to first order in $K/m$, \begin{equation}
\kappa=1+K/(4m).\label{eq:First-order_kappa}\end{equation}
Substituting this in Eqs. (\ref{eq:temp-1}) and (\ref{eq:Flux_Exact}) ,
we finally obtain \begin{equation}
T_{H}=\frac{1}{2\pi}\left[1+\frac{K}{4m}+O\left(\frac{K}{m}\right)^{2}\right]\label{eq:First-order_temp}\end{equation}
for the Hawking temperature, and \begin{equation}
\hat{T}_{\infty}=\frac{K}{4}\left[1+\frac{K}{2m}+O\left(\frac{K}{m}\right)^{2}\right]\label{eq:First-order_flux}\end{equation}
for the outflux.

It is interesting to note that a similar result was previously obtained
for the Hawking outflux emitted from a \emph{non-static, evaporating},
CGHS semiclassical BH. In that case too, the outflux was found to
take precisely the form (\ref{eq:First-order_flux}), this time with
$m$ denoting the Bondi mass (namely the remaining BH mass, as measured
at FNI as a function of $u$). That result, too, was obtained analytically
\cite{Aprox1} and then confirmed numerically \cite{APR,Dori_Ori}.
Here we were able to get this first-order result with a much simpler
calculation (both analytical and numerical), for a static BH model.

\section{Numerical verification}

We carried out a numerical calculation to verify our analytical result
(\ref{eq:First-order_kappa}) for the first-order finite-mass correction
to $\kappa$. To this end we numerically integrated the static semiclassical
evolution equations, namely the nonlinear ODE system (\ref{eq:S-C_Canonic_evo_R},\ref{eq:S-C_Canonic_evo_S}).

The semiclassical CGHS model admits a unique scaling law:
\cite{Aprox,APR} The field
equations remain invariant under a multiplication of both $K$ and
$R$ by the same arbitrary constant (while keeping $\rho$ unchaged).
It follows from this scaling law that the dependence
of $\kappa$ on $m$ and $K$ (in the {\it exact} semiclassical solution) is only through the combination $m/K$.
In our numerical simulations we calculated $\kappa$ as a function
on $m/K$ throughout the range $100<m/K<500$.

\begin{figure}[ht]
\begin{center}
\includegraphics[scale=.55]{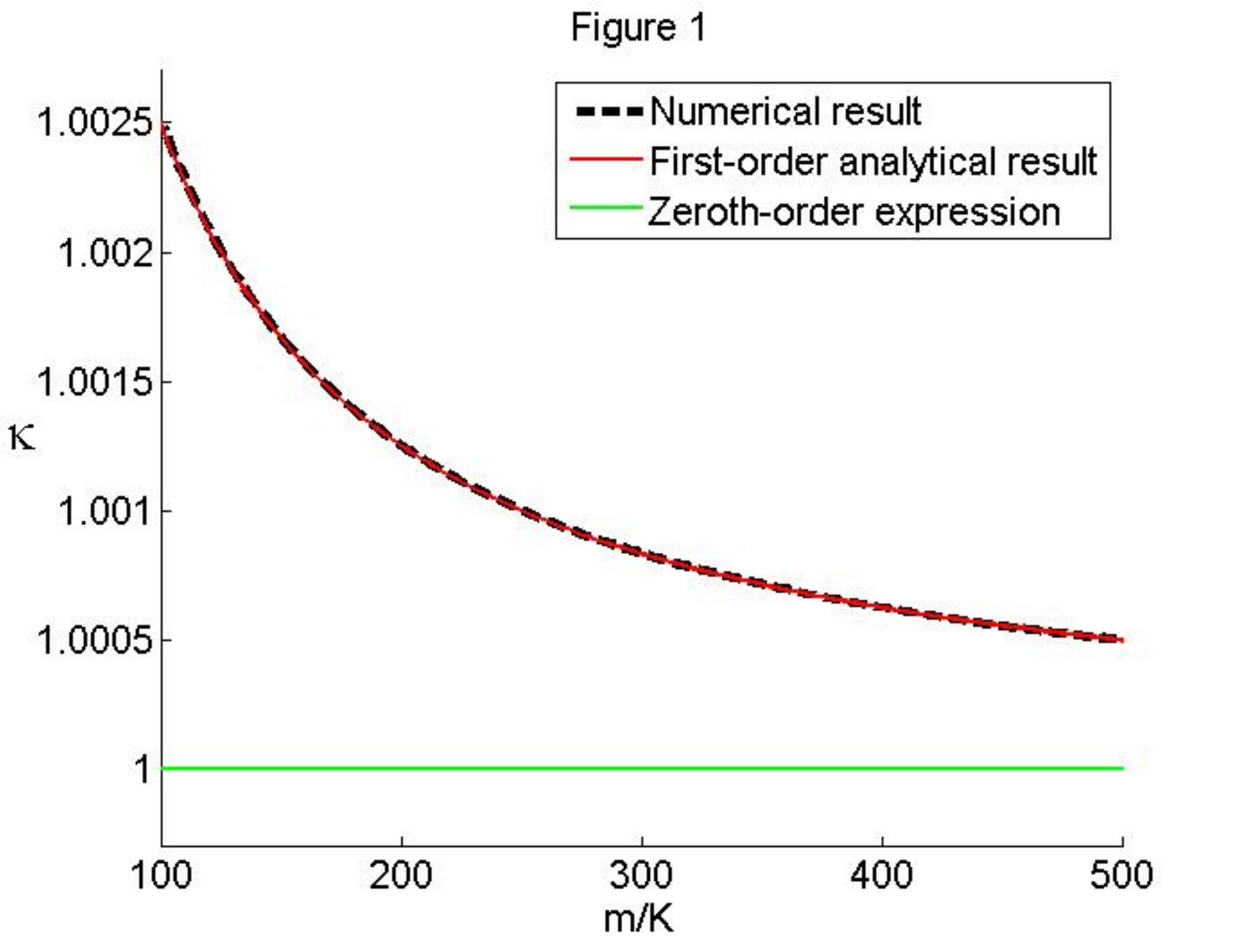}
\caption{ \label{fitting}
The surface gravity $\kappa$ as a function of the BH ``normalized
mass'' $m/K$. The dashed-dotted black curve displays our numerical results,
which were calculated for 250 values of $m/K$ throughout the range
$100<m/K<500$. The first-order theoretical
prediction $\kappa\cong1+k/(4m)$ is the solid red curve. These two
curves are actually visually indistingwishable. For comparison we also display
the corresponding zeroth-order expression $\kappa=1$, the solid green line. }
\end{center}
\end{figure}

In the analytical derivation above we have imposed the regularity
condition $R'_{\left(-\infty\right)}=0$ at the horizon. In addition
we have worked with the gauge in which $S_{\left(\infty\right)}=x,\, S'_{\left(\infty\right)}=1$,
in accord with Eq. (\ref{eq:S-C_RS_inf_asymptotic}). In a numerical
implementation this would mean that we would have to impose boundary
conditions at both boundaries of the simulation, $x\to-\infty$
and $x\to\infty$. This is somewhat inconvenient, though it
can still be done, e.g. by iterations.

However, to simplify the numerical procedure we have chosen a different
approach: We can use a more convenient gauge for the simulation, and
then calculate $\kappa$ using the gauge-invariant expression (\ref{eq:kappa_general-1}).
We have chosen the gauge for the simulation to be $S_{\left(-\infty\right)}=x,\, S'_{\left(-\infty\right)}=1$
at the horizon, and integrated the ODE system (\ref{eq:S-C_Canonic_evo_R},\ref{eq:S-C_Canonic_evo_S})
from the horizon towards infinity. The boundary conditions for $R$
are taken to be $R'_{\left(-\infty\right)}=0$ as well as $R_{\left(-\infty\right)}=m$.
In this way all our boundary conditions are specified at $x=-\infty$.
In this gauge it easily follows from Eq. (\ref{eq:kappa_general-1}) that
$\kappa=1/S'_{\left(\infty\right)}$.

The results of our numerical simulations are presented in Fig. 1. The
graph demonstrates an excellent agreement between the numerical solution
and our theoretical prediction, based on the first-order analysis.
This figure also indicates very clearly that the first-order correction
is really needed: The numerical results deviate substatially from
the zero-order outflux (the horizontal line $\kappa=1$, which is
the infinite-mass limit).

Finally we note that our numerical simulations also reveal a second-order correction term in $\kappa$. This additional term is consistent (within the numerical accuracy) with
$$\frac{3}{32}\left(\frac{K}{m}\right)^{2} . $$
This of course implies a corresponding second-order correction in the Hawking temperature and the outflux.

\section{Discussion}

We have found analytically, and confirmed numerically, that in a large-mass
($m\gg K$), static, CGHS semiclassical BH the surface gravity $\kappa$
admits a small semiclassical correction $\propto K/m$, presented in Eq.
(\ref{eq:First-order_kappa}). Correspondingly, the Hawking temperature
and outflux also get finite-mass corrections, as shown in Eqs. (\ref{eq:First-order_temp},\ref{eq:First-order_flux}).

As was already mentioned above, previous analytical \cite{Aprox1} as well as numerical  \cite{APR,Dori_Ori} analyses of the Hawking radiation
emitted from an \emph{evaporating} CGHS BH revealed the presence of
a semiclassical correction term $\propto K/m$ in the outflux.
Our analysis shows that this correction term is precisely the same
(at order $K/m$) in the static and evaporating cases. Namely, Eq. (\ref{eq:First-order_flux})
holds in the evaporating case as well (provided that one interprets
$m$ as the Bondi mass).

In the evaporating case, one may be tempted to interpret the finite-mass
correction to the outflux as an indication for the non-thermal character
of the Hawking radiation. Such deviations
from thermality might be important for certain aspects of the information puzzle \cite{APR}.
Our analysis of the static case
suggests a different interpretation of this $\propto K/m$ correction
to the outflux: The backreaction of the renormalized stress-energy
tensor $\hat{T}_{\alpha\beta}$ obviously modifies the background
geometry, and as a consequence the BH's surface gravity $\kappa$
is modified too. This inevitably leads to a change in the Hawking temperature,
through the standard relation $T_{H}=\kappa/2\pi$ (which should exactly
hold, for a static BH, even in the semiclassical case).
In turn, this correction in $T_{H}$ naturally
leads to a corresponding correction in the outflux,
through the standard quadratic relation between temperature and outflux
(the 2D analog of the Stefan-Boltzmann law, which states that the outflux
should be proportional to the square of the temperature).
Indeed, this exact quadratic relation between outflux and Hawking temperature is guaranteed to hold in the 2D semiclassical (static) BHs of the CGHS model, as demonstrated in
Eqs. (\ref{eq:temp-1}) and (\ref{eq:Flux_Exact}).
This relation is also reflected in our explicit first-order results, Eqs.
(\ref{eq:First-order_temp}) and (\ref{eq:First-order_flux}).
Thus, in the static case the Hawking outflux remains precisely
thermal (despite the semiclassical backreaction), even though the Hawking
temperature gets a small semiclassical correction.

Turning now to the evaporating case, since the first-order correction to the outflux
precisely agrees with the static result (\ref{eq:First-order_flux}),
it appears that this $O(K/m)$ correction merely reflects the change
in the Hawking temperature due to backreaction (rather than deviations
from thermality). Indeed one may anticipate small deviations from
thermality in the dynamical, evaporating case; however, such deviations
seemingly appear only at second or higher orders in $K/m$.

\section*{Acknowledgment}

This research was supported by the Israel Science Foundation (grant
no. 1346/07)

\appendix

\section{Linear semiclassical correction}

We display here the full solution to the linearized semiclassical
equations (\ref{eq:deltaR''_Explicit},\ref{eq:deltaS''_Explicit}): \[
\delta R=\frac{K}{4}\left[\ln\left(\frac{e^{x}+m}{m}\right)+\frac{e^{x}}{m}\left(x-\ln(e^{x}+m)\right)\right],\]
\[
\delta S=\frac{K}{4m}\left[\frac{m}{e^{x}+m}+x-\ln\left(e^{x}+m\right)\right].\]
The four integration constants were chosen here such that $\delta S$ and $\delta S'$
vanish at infinity (in accord with our gauge condition), and $\delta R,\delta R'$
vanish at the horizon. [By differentiating the last equation
one recovers Eq. (\ref{eq:delta_S',fina_1st order}) for $\delta S'$.]


\end{document}